\documentstyle[aps,multicol,epsf]{revtex}

\renewcommand{\vec}[1]{{\bf #1}}

\newcommand{\be}{\begin{equation}}
\newcommand{\ee}{\end{equation}}
\newcommand{\ber}{\begin{eqnarray}}
\newcommand{\eer}{\end{eqnarray}}

\begin{document}

\title{
  Laplacian growth with separately controlled
noise and anisotropy
}
\author{M.G.~Stepanov${}^{1,2,*}$ and
L.S.~Levitov${}^{3,4,5,\dagger}$}

\address{
  ${}^1$ Institute of Automation and Electrometry,
    Novosibirsk, Russia\\
  ${}^2$ Physics of Complex Systems, Weizmann
    Institute of Science, Rehovot, Israel\\
  ${}^3$ Physics Department, Massachusetts Institute
    of Technology, Cambridge, Massachusetts\\
  ${}^4$ Center for Materials Sciences \& Engineering, 
    Massachusetts Institute of Technology, Cambridge,
    Massachusetts\\
  ${}^5$ Condensed Matter Physics, Weizmann
    Institute of Science, Rehovot, Israel
}

\maketitle

\begin{abstract}
    Conformal mapping models are used to study competition 
of noise and anisotropy in Laplacian growth.
For that, a new family of models is introduced with
the noise level and directional anisotropy controlled 
independently. Fractalization is observed in both 
anisotropic growth and the growth with varying noise.
Fractal dimension is determined from cluster size
scaling with its area. For isotropic growth $d = 1.7$,
both at high and low noise. For anisotropic growth
with reduced noise the dimension can be as low as
$d = 1.5$ and apparently is not universal.
Also, we study fluctuations of particle areas and
observe, in agreement with previous studies, that
exceptionally large particles may appear during the
growth, leading to pathologically irregular clusters.
This difficulty is circumvented by using an acceptance
window for particle areas.

\par\vspace{5mm}\noindent

\end{abstract}


\bigskip\smallskip
\begin{multicols}{2}

\section{Introduction}

In a large class of pattern-forming systems the growth
is controlled by a Laplacian field. In diffusion
limited aggregation (DLA) this field is the
probability density of aggregating
particles~\cite{dla-reviews,WS}. In viscous fingering
it is pressure~\cite{Bensimon-RMP}, and in crystal
growth it can be either a diffusive or a thermal
field~\cite{rmpL80}. After the DLA model was
introduced by Witten and Sander~\cite{WS}, it became
standard to simulate the Laplacian field by random
walkers, which after being released at the periphery
of the system, diffuse towards growing cluster and
freeze on it. To simulate DLA, several numerical
techniques have been developed, of which most powerful
are the off-lattice
algorithms~\cite{jpaBB85,jpaM85,praMBRS87}.

Applications to other Laplacian problems have been
proposed based on the random walks idea.  In
particular, handling problems such as viscous
fingering within this framework requires reducing
noise of individual walkers as well as modeling
surface tension.  Reduction of noise was achieved by
the method of multiple hits~\cite{praT85,jpaSCK85} in
which particles freeze on a particular site adjacent
to the already grown cluster only after this site has
been visited more than $n_{\rm min}$ times, where
$n_{\rm min}$ is an acceptance threshold. The method
of noise reduction~\cite{praT85,jpaSCK85} has been
introduced in the context of the on-lattice DLA
models. More recently, this method was combined with
the off-lattice technique and studied
theoretically~\cite{Eckmann1,Eckmann2,Cafiero}.  In an
advanced version~\cite{Eckmann1} of the multiple hits
method the random walkers move off-lattice and
sticking rules are defined by using a finite number
$m$ of antennas attached to each particle, where, for
instance, $m=4$ for the square lattice DLA.  Each of
$m$ antennas has a counter which scores the number of
times $n$ random walkers arrive on it and is then used
to set a threshold $n_{\rm min}$ for freezing. Having
a finite number of antennas used per each particle
makes the sticking rules in the mutiple hits models
anisotropic.  This anisotropy is essential because it
is significantly amplified by the growth dynamics in
the low noise regime of $n_{\rm min}\gg1$.  This
built-in anisotropy of growth rules has been used to
test universality of DLA~\cite{prlBBRT85} and also to
simulate dendritic crystal
growth~\cite{paB86,Eckmann1}.

It has been proposed that surface tension can be
modeled by introducing a probability $t < 1$ of
freezing upon each encounter with the
cluster~\cite{prbWS83} or by making freezing dependent
on the local neighbors
configuration~\cite{Vicsek84,jpaSCK85,Kadanoff85}.  
In the model~\cite{prbWS83} with $t \ll 1$ each
randomly walking particle freezes only after
encountering the cluster about $t^{-1}\gg1$ times. As
a result, the freezing point is displaced from the
point of first encounter by the distance $d\simeq
t^{-1}$ in the units of particle size. Effectively, in
this model a finite length scale $d$ is introduced
over which the harmonic (Laplacian) measure describing
the probability of the first encounter is being
probabilistically averaged. It has been conjectured in
the works~\cite{Vicsek84,jpaSCK85,Kadanoff85} (and
partially confirmed by various features observed in
the growth patterns) that the length scale $d$
simulates capillary radius in the Laplacian problem
with surface tension.

The field of Laplacian growth, despite being well
developed by now, contains several long-standing
unresolved problems. Firstly, the lattice simulation,
although extremely efficient algorithmically, does not
seem to be a natural starting point for analytic
understanding of large scale phenomena, such as
fingering, fractalization and scaling. Secondly, the
methods of simulating Laplacian growth that have been
used so far are not entirely free of problems, the
most notable being an intrinsic anisotropy of growth
rules. Already the original DLA rules~\cite{WS} use
square lattice and thus are anisotropic. This
anisotropy is weak and reveals itself only in very
large DLA clusters~\cite{jpaBB85,praMBRS87}. However,
when the noise level is reduced using multiple hits,
the underlying lattice anisotropy is
amplified~\cite{Vicsek86,Eckmann2}.  The noise-reduced
growth remains vulnerable to anisotropy even when
off-lattice random walks are used, due to the
anisotropy of freezing rules mentioned above.

It was demonstrated recently that Laplacian growth can
be studied using an entirely different approach based
on iterated conformal maps~\cite{HL}. The
model~\cite{HL} uses analytic functions chosen in such
a way that upon acting on a unit circle they produce
bumps of prescribed size. Iterated $n$ times with the
parameter defining the bump size chosen according to a
certain rule, these maps produce a cluster of $n$
particles of nearly equal size. The conformal model of
growth has become recently a subject of active
work~\cite{Hastings,P1,P2,P3,P4,Somfai}.

The goal of this article is to extend the conformal
mapping methodology to the problems with reduced noise
and growth anisotropy. Here one clear advantage is
that the growth rules using conformal
mapping~\cite{HL} are intrinsically {\it isotropic}.
Because of that one can easily avoid problems
pertinent to other models, in which growth anisotropy
and reduced noise are intertwinned. The main idea
behind the noise reduction method proposed in this
work is to average the Laplacian measure over finite
length which is larger than the particle size in the
original model~\cite{HL}. For that we alter particle
shape and use ``flat'' particles extended along the
cluster boundary and thin in the growth direction. To
compare our method of reducing noise to other
techniques, we note that the positions of flat
particles are chosen strictly according to Laplacian
measure, like in the multiple hits
method~\cite{praT85,jpaSCK85}.  The control over noise
is achieved by suppressing noise at the length scales
shorter than the particle larger dimension. This is in
contrast with the multiple hits method, where noise is
suppressed due to statistical averaging over many
particle growth attempts uniformly over all length
scales down to particle size. Because of the
appearance of a new length scale our method somewhat
resembles the surface tension models used in the DLA
lattice
growth~\cite{prbWS83,Vicsek84,jpaSCK85,Kadanoff85}.

One notable difference from previous models is in the
dependence of the computation time on the achieved
level of noise reduction.  Reducing statistical
fluctuations in the multiple hits models required
increasing the number of random walkers used to grow
the cluster inversely with the noise reduction
parameter. In our method one can reduce noise
arbitrarily without increasing computation length,
simply by varying particle aspect ratio with particle
areas kept fixed. We also demonstrate that growth
anisotropy can be naturally incorporated in the
conformal mapping method without affecting noise
reduction.

Our plan in this article is as follows. We start with
revisiting conformal mapping model. We discuss some
issues ignored before and propose a generalization to
the problems with reduced noise and anisotropy. In
Sec.~\ref{sec-model} we review the model, keeping
focus on aspects that will be important in the rest of
the article. In Sec.~\ref{sec-areas} we study the
distribution of particle areas produced by growth
rules employed in Ref.~\cite{HL}. We observe that
these rules lead to occasional appearance of very
large particles. To fix this problem, we evaluate
particle areas at each growth step and apply an
acceptance criterion for newly grown particles
according to their area. In Sec.~\ref{sec-nonoise} we
describe a model with reduced noise. To suppress noise
we use particles which are thin in the growth
direction and smooth at corners. In
Sec.~\ref{sec-anisotropy} we show how these growth
rules can be generalized for anisotropic growth. In
Sec.~\ref{sec-scaling} we study scaling properties of
all introduced varieties of the model and compare them
with each other. We find that the fractal dimension
estimated from cluster radius scaling is less
sensitive to noise than to anisotropy. For isotropic
growth, both with and without noise reduction, the
dimension is very close to $1.7$. For anisotropic
growth, reducing noise to the level at which
anisotropy reveals itself strongly shifts fractal
dimension to somewhat lower values. In this regime,
the fractal dimension depends on symmetry, and is
found to be $1.62$ for the four-fold symmetry and
$1.5$ for the three-fold symmetry. Finally, in two
appendices we discuss in detail the particles area
distribution and possible improvements of our
numerical procedure.

\section{Conformal mapping model}
\label{sec-model}

We describe growing cluster by a sequence of domains
${\cal D}_0 \subset{\cal D}_1 \subset {\cal D}_2
\subset ...$ corresponding to subsequent growth steps
in time. In the canonical formulation~\cite{WS},
growth occurs due to particles diffusing from infinity
one by one and freezing as soon as they reach the
cluster boundary. The particles concentration $u(\vec
r)$ obeys diffusion equation, which in the
quasi-stationary approximation of slow growth is
written as
  \be\label{bvp}
\nabla^2 u({\vec r}) = 0\ {\rm\ with\ }
u({\vec r}) = \left\{
  \begin{array}{ll}
    0\,, & {\vec r}\in \partial {\cal D}_{n-1}\,; \\
    \frac1{2\pi} \ln|{\vec r}|\,, & |{\vec r}| \to
      \infty\,.
  \end{array}
\right.
  \ee
Zero boundary condition on the cluster ${\cal
D}_{n-1}$ describes freezing of the $n$-th particle
upon arrival with probability one. The points of the
cluster boundary $\partial {\cal D}_{n-1}$ where
subsequent additions are made are selected randomly
with the probability given by the so-called harmonic
measure
  \be\label{dPdl}
dP = |\nabla u| \, dl
\,, \quad
dl\subset \partial {\cal D}_{n-1}
\,,
  \ee
where $dl$ is boundary element of the cluster ${\cal
D}_{n-1}$. As the domain changes, $... \to {\cal D}_n
\to {\cal D}_{n+1} \to ...$, the problem (\ref{bvp})
has to be solved again for every new domain to
determine from (\ref{dPdl})  the new particle position
probability.

A considerable computational simplification of the
problem can be achieved \cite{HL} by using a sequence
of analytic functions $F_n(z)$, $n = 0, 1, 2, ...$, to
represent the domains ${\cal D}_n$. The functions
$F_n$ are chosen so that each of them defines a
conformal one-to-one mapping of the unit disk $|z| \le
1$ on the domain ${\cal D}_n$, including boundary.
Adding a new object to the cluster at the $n$-th
growth step is described by changing the mapping $F_n$
as follows:
  \be\label{F(f)}
F_n(z) = F_{n-1}(f_{\lambda_n,\theta_n}(z))\,, \quad
F_0(z) = z\,.
  \ee
Here the function $f_{\lambda_n,\theta_n}(z)$ maps the
unit circle $|z| = 1$ onto a unit circle with a bump
centered around the point $z = e^{i\theta_n}$ of the
circle. The bump size is determined by the parameter
$\lambda_n$ as discussed below. The angle $\theta_n$
is chosen randomly at each growth step.

The key simplification that arises in the conformal
mapping representation (\ref{F(f)}) is due to the fact
that the harmonic measure (\ref{dPdl}) is translated
into a uniform probability distribution for
$\theta_n$, so that $dP(\theta) = d\theta/2\pi$. Also,
there is no statistical correlation between subsequent
$\theta$'s.

The form of the function $f_{\lambda,\theta}(z)$
growing bumps can be chosen according to the
computational needs \cite{HL}. In this article we use
  \be\label{little-f}
f_{\lambda,\theta}(z) = e^{i\theta} g^{-1} \Big(
\widetilde f_{\lambda} \big( g(e^{-i\theta}z) \big)
\Big)\,,
  \ee
where the function $g(z) = (z-1)/(z+1)$ maps the unit
disk $|z| \le 1$ onto the left half-plane ${\rm Re}\,z
\le 0$, and the function
  \be\label{tilde-f}
\widetilde f_{\lambda}(z) =
h_\lambda(z)/h_\lambda(1)\,,
\quad h_\lambda(z) = z + \sqrt{z^2+\lambda^2}
  \ee
grows a semicircle of radius $\lambda$, as shown in
Fig.~\ref{fig-little-f}.  The function $\widetilde
f_{\lambda}(z)$ is defined in (\ref{tilde-f})  so that
$\widetilde f_{\lambda}(1) = 1$. This ensures that the
mapping (\ref{little-f}) maps $z = \infty$ onto
itself.

\begin{minipage}{3.1in}
\begin{figure}
  \begin{center}
    \begin{picture}(216,146)
      \epsfxsize=3in
      \put(0,-5){\epsfbox{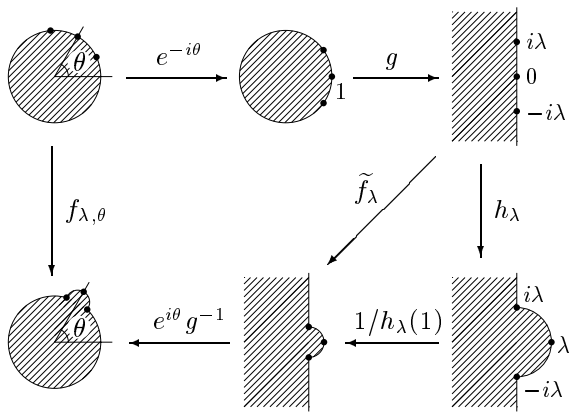}}
    \end{picture}
  \end{center}
  \caption{The sequence of mappings constituting
    $f_{\lambda,\theta}(z)$ defined by
    (\protect\ref{little-f}), (\protect\ref{tilde-f}).
    \label{fig-little-f}}
\end{figure}
\vspace{0mm}
\end{minipage}

Ideally, the values $\lambda_n$ defining particle size
should be chosen so that all particle areas are equal.
In the conformal mapping model \cite{HL} this is
approximately realized via predicting the bump size to
be obtained at the $n$-th step using the Jacobian of
the already-grown cluster mapping $F_{n-1}$.

The argument is as follows. The area of the
semicircular bump grown using $\widetilde
f_{\lambda_n}(z)$ is $\pi\lambda^2_n (1 +
O(\lambda_n^2))/8$. The area of the corresponding bump
produced by $F_n(z)$, at small $\lambda_n$, is
approximately $|J_{n-1}|^2 \pi\lambda^2_n/2$, where
$J_{n-1}$ is the Jacobian of $F_{n-1}$ evaluated at
the position of the $n$-th bump:
  \be\label{Jacobian}
J_{n-1} = F'_{n-1}(z = e^{i\theta_n})\,.
  \ee
Hence, to compensate for stretching due to the
Jacobian $J_{n-1}$, one has to choose the values of
$\lambda_n$ as follows:
  \be\label{bump-size}
\lambda_n = |J_{n-1}|^{-1}\lambda_0\,,
  \ee
where the parameter $\lambda_0$ defines particle size.
For the growth involving particles of very small size
the rule (\ref{bump-size})  would have been sufficient
to ensure identical areas of all bumps.  For our
problem, in which the bump sizes are small but finite,
the areas are only approximately equal. However, one
can demonstrate that, after certain improvements
discussed in Section~\ref{sec-areas}, the rule
(\ref{bump-size}) produces bumps with sufficiently
close areas.

The form (\ref{little-f}), (\ref{tilde-f}) of the
mapping $\widetilde f$ has several nice features.
First, since fractional linear function $g(z)$ maps a
circle onto a circle, the mapping $f_{\lambda,\theta}$
produces a crescent-shaped particle with circular
boundary (see Fig.~\ref{fig-little-f}). Second, a
simple calculation shows that the particle curvature
radius equals $\lambda$. The latter has the following
consequence. Consider growth starting from a circular
cluster of radius $r$, described by the mapping $F(z)
= rz$. The mapping $F(f_{\lambda,\theta}(z))$ then
produces a particle of curvature radius equal to
$\lambda r$.  After the value of $\lambda$ is chosen
according to the rule (\ref{bump-size}), $\lambda =
\lambda_0/F' = \lambda_0/r$, the particle radius
becomes equal to $\lambda_0$, independently of the
cluster radius $r$. The area of this particle is
readily evaluated:
  \ber\label{a-star}
a(\lambda_0,r) & = & a_\ast+\lambda_0
r-(r^2-\lambda_0^2) \tan^{-1}(\lambda_0/r)\,,
\nonumber\\
& &a_\ast = \pi\lambda_0^2/2\,.
  \eer
The area $a(\lambda_0,r)$ varies between $a_\ast$ for
$\lambda_0\ll r$ and $2a_\ast$ for $\lambda_0\gg r$.
For a generic non-circular cluster the particle area
cannot be found analytically. Statistics of the areas
will be discussed in Section~\ref{sec-areas}. We will
see that typical area of a particle is of the order of
$a_\ast$.

The overall size of the cluster ${\cal D}_n$
grown according to (\ref{F(f)}), (\ref{bump-size})
is well characterized~\cite{HL} by the mapping $F_n$
stretching factor at large scales:
  \be\label{R-definition}
R_n = F'_n(z\to\infty) =
\prod\limits_{k=1}^n
f'_{\lambda_k,\theta_k}(z\to\infty)
\,.
  \ee
The cluster radius $R_n$ can be conveniently evaluated
using (\ref{R-definition}) together with the property
$g(\infty) = 1$ as follows:
  \be\label{R-p=1}
R_n = \prod\limits_{k=1}^n \left( \widetilde
f'_{\lambda_k,\theta_k}(z = 1)\right)^{-1} =
\prod\limits_{k=1}^n \left(1+\lambda_k^2
\right)^{1/2}\,.
  \ee
The reason for $R_n$ to be an accurate measure of the
cluster ${\cal D}_n$ radius lies in the properties of
so-called univalent functions~\cite{HL,P1}.

At large $n$ the cluster radius $R_n$ is expected to
grow as $n^\alpha$, where $\alpha$ is a numerical
constant. This is consistent with (\ref{R-p=1})
provided that $\lambda_n^2n \to 2\alpha$ at large $n$
\cite{HL,Halsey-harmonic}. The growth problem
(\ref{bvp}), (\ref{dPdl}) is believed to give rise to
fractal objects with fractal dimension $d < 2$.  
There are several conventional definitions of fractal
dimension of a growing cluster~\cite{dla-reviews}. In
this article we employ scaling of the cluster size
with its area. Also, one can use box counting, or the
relation between average growth velocity in a strip
and the strip width.

Taking $R_n$ defined in (\ref{R-definition}) as a
cluster radius provides a numerically efficient method
for calculating fractal dimension. For that one looks
for a scaling relation of the form
  \be\label{R-scaling}
R_n \propto A_n^{1/d}\,,
  \ee
where $A_n$ is total area of the cluster ${\cal D}_n$.
The dimension $d$ is related with the parameter
$\alpha$ describing scaling of $\lambda_n$ as follows:
$d = \alpha^{-1}$, which is true provided $A_n \propto
n$. In our simulation we make sure that individual
particle areas have a sufficiently narrow distribution
(see Section~\ref{sec-areas}), and thus total cluster
area is indeed proportional to the particle number.

Scaling properties of the growth problem described
above have been explored by several
groups~\cite{HL,Hastings,P1,P2,P3,P4,Somfai}. It was
concluded that the properties of the growth resulting
from the conformal mapping model match those of the
lattice DLA models. Below we revisit the relation
between the problem (\ref{bvp}), (\ref{dPdl}) and the
conformal mapping model (\ref{F(f)}),
(\ref{little-f}), (\ref{bump-size}) and discuss
several interesting extensions of this model.

\section{Area distribution}
\label{sec-areas}

In the original work \cite{HL} it was assumed that the
rule (\ref{bump-size}) is sufficient to produce
particles with nearly equal areas. This assumption was
apparently consistent with the cluster images in which
each particle is represented by one or few points.
To investigate this issue more
closely, in this work we have chosen a different
method of representing particles, in which the exact
boundary of each particle is shown. An example of a
cluster with the boundaries of all individual
particles displayed (see Fig.~\ref{lap}), demonstrates
that the areas of almost all of the particles are
indeed quite close. However, there is also a number of
exceptional particles of large areas.

Large particles tend to appear within fjords and seal
the space between well developed branches. Typically,
this happens when particle growth is attempted on the
periphery of an actively growing region. Insufficiency
of the rule (\ref{bump-size}) for keeping particle
areas small is caused by fluctuations of the Jacobian
$F'_n(z)$ over the unit circle $|z| = 1$. These
fluctuations can be large in magnitude and also very
abrupt, happening on the scale of the order of
$\lambda_n$ within the circle arc mapped onto the
particle boundary. In the case when a newly grown
particle overlaps with such a fluctuation, it can be
``artificially stretched'' under the mapping.

The appearance of large particles has been reported in
Ref.~\cite{P1} and a method for eliminating them was
proposed, based on choosing an optimal shape of
particles produced by the mapping
$f_{\lambda,\theta}(z)$. It was argued~\cite{P1} that
the best value of the parameter $0<a<1$ in the mapping
defined in Ref.~\cite{HL} is given by $a=2/3$. This
value provides a compromise between abundance of large
particles at $a\to 0$ and needle-like particles shape
at $a\to 1$. Since in this article the particle shape
will be used to tune noise, we employ a different
method for eliminating large particles, as described
below.

\begin{minipage}{3.1in}
\begin{figure}
  \begin{center}
    \begin{picture}(216,163)
      \epsfxsize=3in
      \put(0,-5){\epsfbox{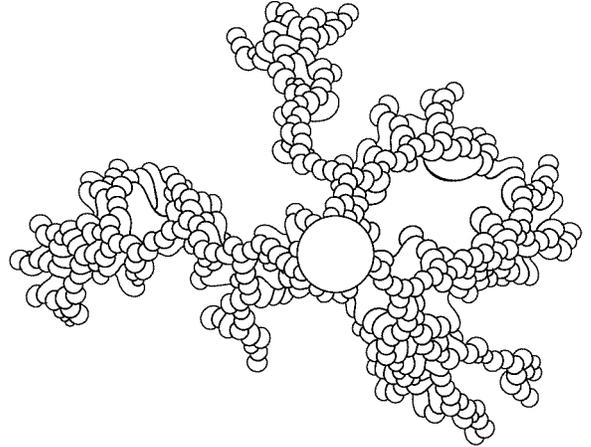}}
    \end{picture}
  \end{center}
  \caption{Cluster of $N = 400$ particles grown using
    the model (\protect\ref{F(f)}),
    (\protect\ref{bump-size}) with $\lambda_0 = 0.2$
    and  $f_{\lambda,\theta}$ of the form
    (\protect\ref{little-f}), (\protect\ref{tilde-f}).
    The boundary of each particle is displayed. Note
    large particles which appear rarely and seal
    fjords. \label{lap}}
\end{figure}
\vspace{0mm}
\end{minipage}

To study the role of exceptional particles, we
calculate the particle areas generated by the growth
model (\ref{F(f)}), (\ref{bump-size}).  The method
employed to evaluate particle areas is the following.
For the particle grown at a step $n$, we start with
few points on the unit circle $|z| = 1$ which are
mapped by $F_n$ on the particle boundary.
Subsequently, we add new points on the circle $|z| =
1$ in between the old points, and compute distances
between images of neighboring points under {\it all}
mappings
  \be
f_n\,, \quad f_{n-1}\circ f_n\,, \quad ... \quad ,
\quad F_n = f_1 \circ f_2 \circ ... \circ f_n\,,
  \ee
where $f_k$ is a shorthand notation for
$f_{\lambda_k,\theta_k}$ and $\circ$ stands for
mapping composition. We keep adding new points until
the distances between the images of all neighbors will
not exceed $\gamma \lambda_0$, where $\gamma\ll1$ is a
numerical factor. We used the above procedure with
$\gamma = 0.05$, which produces about two to four
hundred points per particle.

This method enables one to have an accurate graphical
representation of each particle, as demonstrated in
Fig.~\ref{lap}, and also to evaluate particle areas
with the accuracy on the level of $0.1\%$. This is
illustrated in Fig.~\ref{argrow}(a) showing how the
cluster total area $A_n$ is changing during the growth
of the cluster displayed in Fig.~\ref{lap}. The area
$A_n$ grows as a function of $n$ in small steps of
order $a_\ast = \pi\lambda_0^2/2$, alternated with
occasional jumps of a much larger magnitude. The jumps
correspond to the appearance of large particles which
seal inner cluster regions. The decomposition of the
growth of $A_n$ into the smooth and singular parts is
revealed more clearly in Fig.~\ref{argrow}(b) showing
the dependence of $A_n$ versus $n$ for the same growth
as in Fig.~\ref{argrow}(a), with a linear function
$2.1 \, a_\ast n$ subtracted.

\begin{minipage}{3.1in}
\begin{figure}
  \begin{center}
    \begin{picture}(216,137)
      \epsfxsize=3in
      \put(0,-5){\epsfbox{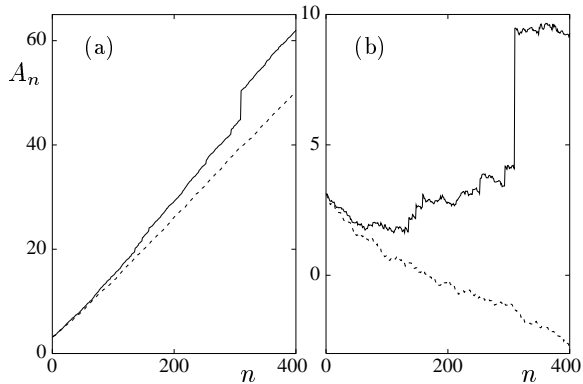}}
    \end{picture}
  \end{center}
  \caption{(a) {\it Solid line\/}: total cluster area
    $A_n$ dynamics for the growth of the cluster in
    Fig.~{\protect\ref{lap}}, $N = 400$, $\lambda_0 =
    0.2$; {\it Dotted line\/}: the area $A_n$ versus
    $n$ for the growth in which only particles with
    the area inside the window $[0,3a_\ast]$ are
    accepted, other growth parameters are the same;
    \hskip 3in
    \hbox{\hskip 5pt} (b) The same area dynamics as in
    (a) with a linear function subtracted: $A_n - 2.1
    \, a_\ast n$ versus $n$. \label{argrow}}
\end{figure}
\vspace{0mm}
\end{minipage}

A histogram of the individual particles areas $a_n =
A_n - A_{n-1}$ is plotted in Fig.~\ref{arpdf}(a). The
area distribution was obtained by averaging over $10$
realizations of the first $1000$ growth steps with the
parameter $\lambda_0 = 0.2$ (same as in
Figs.~\ref{lap},~\ref{argrow}).  The area distribution
${\cal P}(a_n)$ is sharply peaked about $2.1 \,
a_\ast$ and has a number of other interesting
properties that will be discussed in
Appendix~\ref{appen-areas}.

The feature of the distribution ${\cal P}(a_n)$
displayed in Fig.~\ref{arpdf}(a) which corresponds to
the exceptionally large particles is the tail
stretching far to the right from the main peak. Note
that only the beginning of the tail is displayed in
Fig.~\ref{arpdf}(a) because the weight of the tail in
the probability distribution is insignificant, and so
the values of ${\cal P}(a_n)$ far in the tail are too
small to be visible on the scale of the peak.

To display the tail we replot the distribution ${\cal
P}(a)$ on a log--log scale, as shown in
Fig~\ref{artail}. The right tail of ${\cal P}(a)$ is
power-like, ${\cal P}(a)\propto a^{-\mu}$ with $\mu
\approx 2.5$. Since $\mu > 2$, the first moment
$\langle a\rangle = \int a {\cal P}(a)\,da$ is finite
(and thus $\langle a\rangle \simeq a_\ast$). However,
since $\mu < 3$, the second moment of ${\cal P}(a)$ is
divergent. The existence of the mean particle area
$\langle a\rangle$ means that $\langle A_n\rangle
\propto n$. However, the absence of variance implies
that the fluctuations of $A_n$ about the mean value
are non-Gaussian and larger than required by the
central limit theorem.
Both features are clear in the sample $A_n$ dependence
in Fig.~\ref{argrow}.

Let us remark that the tail in Fig.~\ref{artail} in
its far end is apparently somewhat steeper than
$a^{-\mu}$. We believe that this deviation from the
$a^{-\mu}$ behavior is due to finite number $N =
2000$ of time steps in the growth samples
used to obtain ${\cal P}(a)$. In a finite cluster
there is an upper cutoff on possible particle areas.
This makes the far tail of ${\cal P}(a)$
non-stationary, shifting the cutoff to larger areas as
$N$ increases.

\begin{minipage}{3.1in}
\begin{figure}
  \begin{center}
    \begin{picture}(216,141)
      \epsfxsize=3in
      \put(0,-5){\epsfbox{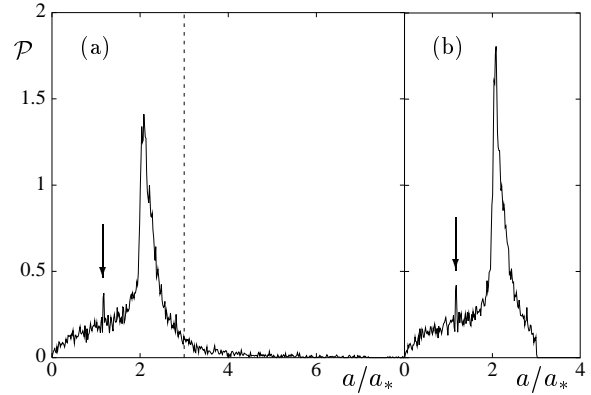}}
    \end{picture}
  \end{center}
  \caption{(a) Probability distribution function of
    particle areas normalized by $a_\ast = \pi
    \lambda_0^2/2 \approx 0.063$ (see
    (\protect\ref{a-star})). Statistics was taken over
    $10$ independent runs of the growth with $N =
    1000$, $\lambda_0 = 0.2$. Note the peak at $\sim
    2.1 \, a_\ast$ and the tail corresponding to
    exceptionally large particles. The largest area
    observed was $\approx 87.8 \, a_\ast$ (see
    Fig.~\protect\ref{lap}). Small peak marked by
    arrow is due to the primary particles growing
    directly on the unit circle.
    \hskip 3in
    \hbox{\hskip 5pt} (b) Same as in (a) for the
    growth in which only particles with areas in the
    window $[0,3a_\ast]$ are accepted. \label{arpdf}}
\end{figure}
\vspace{-5mm}
\end{minipage}

\begin{minipage}{3.1in}
\begin{figure}
  \begin{center}
    \begin{picture}(216,159)
      \epsfxsize=3in
      \put(0,-6){\epsfbox{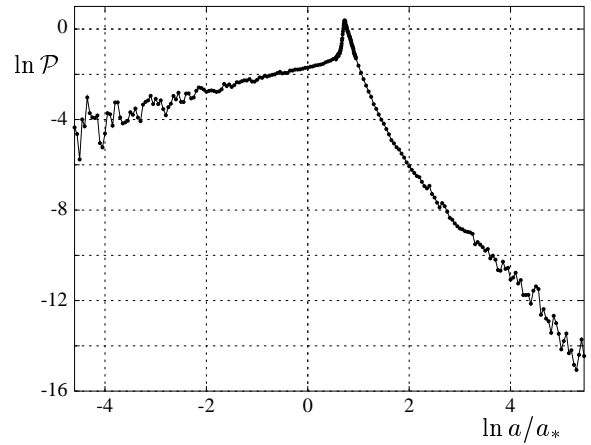}}
    \end{picture}
  \end{center}
  \caption{Logarithm of the probability distribution
    function of particle area $\ln {\cal
    P}(a/a_\ast)$. Calculated from 83 realizations
    with $N = 2000$, $\lambda_0 = 0.8$.
    \label{artail}}
\end{figure}
\vspace{0mm}
\end{minipage}

 
Clearly, one would like to inhibit the appearance of
large particles with areas in the tail of ${\cal
P}(a)$.  This is desirable because, even though the
tail is quite thin and large particles are rare,
occasionally appearing extremely large particles may
affect macroscopic characteristics of the growth. In
particular, the relatively slow power law decrease in
the tail, ${\cal P}(a)\propto a^{-\mu}$, may affect
scaling of the cluster size $R_n$ and/or the numerical
accuracy of the scaling exponent. To eliminate the
growth of large particles, we choose an acceptance
threshold $a_{\rm max} = 3a_\ast$ to truncate the tail
of the distribution in Fig.~\ref{arpdf}(a). Then, for
each growth step, we calculate the new particle area
$a_n$. The particle is accepted only if $a_n \le
a_{\rm max}$, otherwise the particle is discarded and
a new attempt of particle growth is made. An example
of the cluster grown according to these rules is
displayed in Fig.~\ref{fig-p=1}.

One can notice immediately that the overall structure
of the branches in Fig.~\ref{fig-p=1} is much more
regular than that in Fig.~\ref{lap}. The distribution
of areas for such a growth is shown in
Fig.~\ref{arpdf}(b). Within the acceptance window $[0,
a_{\rm max}]$ the distribution ${\cal P}(a)$ repeats
in all details the distribution shown in
Fig.~\ref{arpdf}(a) for the growth with all particle
areas accepted.

\begin{minipage}{3.1in}
\begin{figure}
  \begin{center}
    \begin{picture}(216,140)
      \epsfxsize=3in
      \put(0,-5){\epsfbox{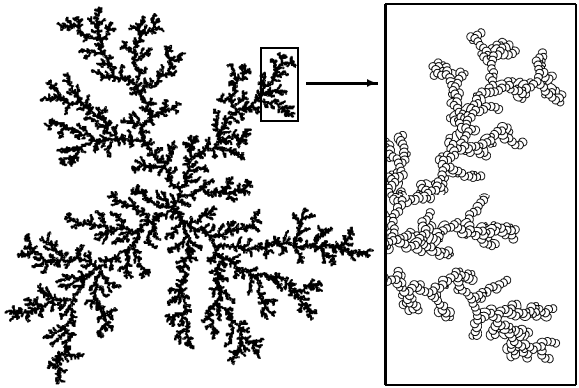}}
    \end{picture}
  \end{center}
  \caption{Cluster grown with the particle area
    acceptance window $[0,3a_\ast]$ (see
    Fig.~\protect\ref{arpdf}(b)), $N = 17545$,
    $\lambda_0 = 0.8$. \label{fig-p=1}}
\end{figure}
\vspace{0mm}
\end{minipage}

For the growth with large particles eliminated, the
area $A_n$ increases as a linear function of the step
number. Average increment of $A_n$ is given by the
mean value of $a_n$ taken from the distribution shown
in Fig.~\ref{arpdf}(b). To verify this, we plot $A_n$
versus $n$ in Fig.~\ref{argrow} for the same growth
parameters as those used in Fig.~\ref{lap}, where 
only the areas up to $3a_\ast$ are accepted. Note a small
difference between the slope of the dependence at
$n \le 50$ and at larger $n$ that appears because of
relatively smaller size of the primary particles
growing directly on the unit circle.

The growth model augmented with the area acceptance
criterion has a new parameter $a_{\rm max}/a_\ast$. In
principle, choosing different values of $a_{\rm max}$
gives rise to different growth patterns. However, as
long as the window $[0,a_{\rm max}]$ contains much of
the ${\cal P}(a)$ peak area, we do not observe any
qualitative change in the growth.

Scaling properties of the growth can be studied in
several ways. Previous studies of
scaling~\cite{HL,P1,P2,P3,P4} are based on the
relation $R_n\propto n^\alpha$, where $R_n$ was
obtained for the growth with unrestricted areas. 
However, it
would be more in the spirit of the notion of a fractal
to use the relation (\ref{R-scaling})
between cluster size and its area. This clearly would
not work well with large particles being present, because
statistical fluctuations
of the cluster area $A_n$ are quite large
in this case.
On the other hand, for the
growth with restricted areas used in this
work, the fluctuations of $A_n$ are
reduced to the level consistent with the central
limit theorem, and thus one can employ the relation
(\ref{R-scaling}) to study scaling.


Of course, it is not clear {\it a priori} whether the
growth with area cutoff is equivalent macroscopically
to the growth with unrestricted areas. On general
grounds, one may expect the growth to be significantly
altered by eliminating large particles. Whether this
is true can be indirectly tested by comparing the
$R_n$~vs.~$n$ dependence for the growth with
unrestricted areas with the fractal dimension obtained
from the relation (\ref{R-scaling}) for the growth
with area cutoff. We note in that regard that the
scaling exponent $d = 1.7$ found below (see
Section~\ref{sec-scaling}) matches exactly the value
found in Ref.~\cite{HL}. However, although the
presence or absence of large particles seems to be
irrelevant for the cluster size scaling, other growth
characteristics, such as the structure of branches and
fjords, are likely to be more sensitive to the method
of treating large particles.

We postpone the discussion of various details and
features of the area distribution ${\cal P}(a)$ to
Appendix~\ref{appen-areas}. In the remaining part of
the article we use the conformal mapping model
augmented with the area acceptance criterion to study
several interesting Laplacian growth problems.

\section{Noise-reduced Laplacian growth}
\label{sec-nonoise}

Roughness of the growing cluster is mainly due to two
factors: shot noise and the Mullins-Sekerka
instability \cite{Mullins-Sekerka}.  The shot noise
results from the randomness of the aggregating
particles positions, and so it contributes to the
fluctuations equally on all spatial scales down to the
particle size. The Mullins-Sekerka instability is due
to aggregation rate enhancement near the tips, which
leads to incremental growth of perturbations of a smooth
front. The wavenumber dependence of the growth rate for a
harmonic modulation of an interface moving with
average velocity $v$ is given by $\gamma_k=v |k|$.
The linear
$k$-dependence of $\gamma_k$ implies that the
instability develops first on the smallest scale, in
our problem given by the particle size.

To study the ultraviolet cutoff role, i.e., the effect
of short distances on the noise and the instability,
it is of interest to introduce a parameter in the
problem which allows to shift the value of the cutoff
scale to values larger than the particle size. One
expects that upon doing so both the noise and the
instability growth rate will be reduced.

In the mapping model, the noise level can be controlled
by altering the shape of aggregating particles. Below
we show how by changing the function $\widetilde
f_\lambda(z)$, defined by (\ref{tilde-f}), one can
create ``flat'' particles which are wide along the
interface and thin in the growth direction. The reason
that noise is suppressed due to using flat particles
is the following. In this growth, a particular
displacement of the growing cluster boundary amounts 
to a larger number of layers than in the case of 
rounded particles used in Ref.~\cite{HL}. 
Then, due to statistical averaging
over many particle layers the boundary displacement
becomes less erratic, and so the noise is reduced.
Quantitatively, the noise suppression factor can be
estimated as a square root of the particle aspect
ratio.

Flat particles can be produced by modifying
$\widetilde f_\lambda(z)$ as follows:
  \ber\label{new-tilde-f}
\widetilde f_{\lambda,p}(z) & = & W h_{\lambda_p}^{-1}
\left( {\textstyle\frac1p} h_{\lambda_p} \big(
h_\lambda(z) \big) \right)
\,, \quad
\lambda_p = \frac{2\lambda}{p+1/p}
\,, \\
&& W = \left(h_{\lambda_p}^{-1} \big(
{\textstyle\frac1p} h_{\lambda_p}(h_\lambda(1))
\big)\right)^{-1}\,, \quad p \ge 1\,.
  \eer
The function $h_\lambda(z)$ is defined in
(\ref{tilde-f}), and its inverse has the form
$h_\lambda^{-1}(z) = \frac12(z-\lambda^2/z)$. The
factor $W$ is introduced in order to have $\widetilde
f_{\lambda,p}(1) = 1$, like for the function
$\widetilde f_\lambda(z)$ defined by (\ref{tilde-f})
above.  The resulting function (\ref{little-f})
satisfies $f_{\lambda,\theta}(\infty)=\infty$, which
ensures the property $F_n(\infty) = \infty$ for all
$n$.

The mapping produced by the function
(\ref{new-tilde-f}) is illustrated in
Fig.~\ref{fig-tilde-f}.  Note that because of
$h_{\lambda_p}(i\lambda)/p = i\lambda_p$, the square
root singularities in $\widetilde f_{\lambda,p}$ at
$z=\pm i\lambda$ are absent for all $p>1$. Instead,
the mapping composition (\ref{new-tilde-f}) produces
weaker singularities of the form $(z\pm
i\lambda)^{3/2}$. This smoothens the corners of the
particles, as shown in Fig.~\ref{fig-tilde-f}.

\begin{minipage}{3.1in}
\begin{figure}
  \begin{center}
    \begin{picture}(216,146)
      \epsfxsize=3in
      \put(0,-5){\epsfbox{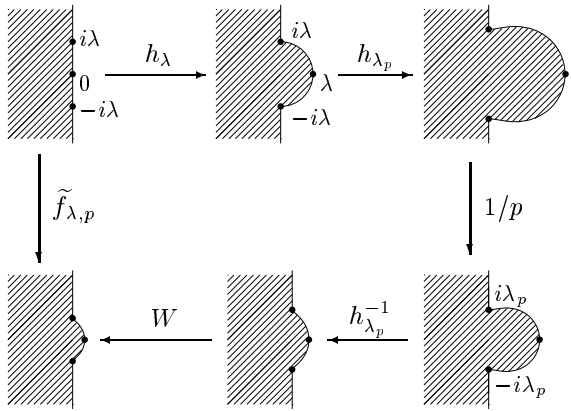}}
    \end{picture}
  \end{center}
  \caption{The sequence of mappings constituting
    $\widetilde f_{\lambda,p}(z)$ as defined by
    (\protect\ref{new-tilde-f}), $p = 1.5$.
    \label{fig-tilde-f}}
\end{figure}
\vspace{0mm}
\end{minipage}

Qualitatively, under variation of $p$ the particle
shape evolves as follows. At $p = 1$ the mapping
$\widetilde f_{\lambda,p}$ form (\ref{new-tilde-f})
coincides with (\ref{tilde-f}). Increasing $p$
produces particles with growing aspect ratio, as can
be seen from comparing the zoom parts of
Figs.~\ref{p1a4},~\ref{p2a4}~and~\ref{p3i}.

To illustrate the effect of $p$ on the particle shape,
consider the mapping function (\ref{new-tilde-f}) in
the limit $p\gg1$. First, one can rewrite
(\ref{new-tilde-f}) as
  \be\label{another-form}
\widetilde f_{\lambda,p}(z)=W'\left(h_\lambda(z)
-\frac{\lambda_p^2(p^2-1)}{2h_{\lambda_p}(h_\lambda(z))}
\right)
\,,
  \ee
where $W'$ is a prefactor chosen so that $\widetilde
f_{\lambda,p}(1)=1$. Expanding (\ref{another-form}) to
lowest order in $1/p$, one obtains
  \be\label{p-expansion}
\widetilde f_{\lambda,p}(z) = z +
\frac{\lambda^2}{p^2}\left(
\frac{z+2\sqrt{z^2+\lambda^2}} 
{(z+\sqrt{z^2+\lambda^2})^2} - Cz
\right)\,,
  \ee
where $C = (2h_\lambda(1)-1)/h_\lambda^2(1)$ and
$h_{\lambda}$ is defined in (\ref{tilde-f}). The
boundary of the particle produced by $\widetilde
f_{\lambda,p}(z)$ of the form (\ref{p-expansion}), to
lowest order in $1/p$, is
  \be
x = \frac{2}{\lambda^2 p^2} \left( \lambda^2 - y^2
\right)^{3/2}\,,
  \ee
where $x+iy = z$. The area of this particle is
$3\pi\lambda^2/4p^2$. Mapped by $g^{-1}$, according to
(\ref{little-f}), the area is multiplied by a factor
equal to $4$ at $\lambda\ll1$.

One can use the growth mapping model (\ref{F(f)}),
(\ref{bump-size}) with the new function
(\ref{new-tilde-f}) to grow clusters in a pretty much
the same way as it was done for the model with $p=1$
in Section~\ref{sec-areas}. The first step is to study
the particle areas distribution for the growth with
unrestricted areas. The distribution looks similar to
that in Fig.~\ref{arpdf}, containing central peak and
the tails corresponding to very large and very small
particles.  In this case the peak is somewhat wider
than for the $p=1$ case. However, much of its weight
in the distribution ${\cal P}(a)$ is still contained
in the window $[0,3a_\ast]$. Here the ``standard
area'' $a_\ast$ is defined, by analogy with
(\ref{a-star}), as the area of a particle grown over a
perfectly flat interface. (For $p\ne 1$ there is no
closed form expression for the particle area, like
(\ref{a-star}), and so one has to calculate $a_\ast$
numerically.)

As before, at each growth step we choose $\theta_n$
randomly, $0 \le \theta_n < 2\pi$, and calculate the
parameter $\lambda_n$ using (\ref{bump-size}), i.e.,
based on the particle area predicted from the Jacobian
$J_{n-1}$. Then we evaluate the actual area $a_n$ of
the particle. To inhibit the appearance of large
particles, we use the acceptance window $[0,3a_\ast]$.
If $a_n > 3a_\ast$, the particle is not accepted and a
new growth attempt is made.

\begin{minipage}{3.1in}
\begin{figure}
  \begin{center}
    \begin{picture}(216,139)
      \epsfxsize=3in
      \put(0,-5){\epsfbox{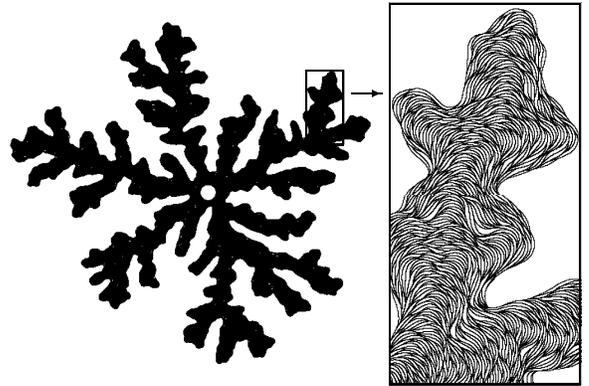}}
    \end{picture}
  \end{center}
  \caption{Cluster grown with $p=3$, $\lambda_0 =
    0.2$. The number of growth steps $N = 15043$.
    \label{p3i}}
\end{figure}
\vspace{0mm}
\end{minipage}

An example of growth with $p=3$ and $\lambda_0 = 0.2$
is displayed in Fig.~\ref{p3i}. In the inset we zoom
on the details of one finger.  Note that individual
particles are indeed quite flat and are evenly spread
over the cluster boundary, indicating reduced noise.
The growing interface is overall very smooth, without
sharp tips or corners. Also, the fingers are much
thicker than for the $p=1$ growth (see.
Fig.~\ref{fig-p=1}).

The cluster size $R_n$ is defined by
(\ref{R-definition}).  As in Section~\ref{sec-areas},
the terms in the product (\ref{R-definition}) can be
evaluated using the relation
$f'_{\lambda_p,\theta}(\infty)= 1/{\widetilde
f}'_{\lambda,p}(1)$, where
  \be\label{new-f'}
\widetilde f_{\lambda,p}'(1) =
\frac{h_\lambda(1)/\sqrt{1+\lambda^2}}
{\sqrt{h^2_\lambda(1) + \lambda_p^2}}
\frac{h_{\lambda_p}^2(h_\lambda(1)) + p^2
\lambda_p^2}{h_{\lambda_p}^2(h_\lambda(1)) - p^2
\lambda_p^2}\,.
  \ee
In the following Section~\ref{sec-scaling} we use
(\ref{new-f'}) along with (\ref{R-definition}) to
evaluate the cluster radius $R_n$ and study its
scaling.





The appearance of the cluster in Fig.~\ref{p3i} shows
that using flat particles indeed helps to reduce
statistical fluctuations. In this model, effective
averaging of harmonic measure is due to the presence
of a tangential-to-boundary length scale set by
particle larger dimension.  This length scale is
controlled by the parameter $p$ and becomes large at
$p\gg1$, if measured in the units of particle size
$\sqrt{a_\ast}$.  Noise reduction takes place due to
the absence of fluctuations with wavelength smaller
than the particle larger dimension, resulting in the
shift of the shot noise spectrum cutoff wavenumber
from $2\pi/\sqrt{a_\ast}$ to lower values as the
parameter $p$ is increased.  Because of reduced noise,
as compared to the $p=1$ case, more agregation events
of flat particles are needed to reach a given radius
of the cluster.


Averaging over a tangential length scale is somewhat
similar to that used in the on-lattice DLA models to
simulate surface
tension~\cite{prbWS83,Vicsek84,jpaSCK85,Kadanoff85}.  
In these works freezing of random walkers upon each
ecounter with the cluster was described by a finite
probability $t<1$ which could be a function of
occupancy of the sites around freezing point.  Since
freezing of each particle typically takes place after
about $t^{-1}$ encounters with the cluster, at $t\ll1$
these models are characterized by a large length scale
over which Laplacian measure is probabilistically
averaged. Similarly, the flat particles used in our
model can be thought of as a result of averaging over
possible particle positions within a finite length
scale taken over harmonic measure. Moreover, there is
a slight dependence of particle size on their growth
position: the particles appearing near the tips are
somewhat smaller than those appearing in the concave
regions (see Fig.~\ref{p3i}). This correlation is
consistent with the surface tension interpretation.


The crucial difference, however, is that particle
positions in our model are chosen according to the
unaltered harmonic measure, whereas in the surface
tension models particle freezing depends on local
boundary geometry.  From that point of view our model
is more similar to the multiple hits
models~\cite{praT85,jpaSCK85} in which statistical
averaging of the harmonic measure over particle growth
attempts is used to control noise. In these models
noise reduction is achieved by averaging over
independent random walkers with a threshold on the
minimal number of visits of each site required before
freezing at this site. Since independent walkers
arrive at very distant points of cluster boundary,
this averaging is not characterized by an additional
large length scale and thus bears no resemblance to
surface tension.
 

The models using finite freezing probability $t<1$
have been shown to give rise to clusters with thick
branches. The Laplacian character of the dynamics and
the analogy of the averaging length scale with
capillary radius was pointed out~\cite{prbWS83} and a
relation with the Saffman--Taylor problem with surface
tension has been conjectured~\cite{Kadanoff85}.  
Because of the large length scale appearing in our
averaging scheme, here a similar relation to the
problems with surface tension can be conjectured.
Indeed, the growth displayed in Fig.~\ref{p3i} looks
like a typical fingering pattern observed in the
Saffman-Taylor problem with surface tension.  As a
word of caution, one should realize that all available
evidence for the equivalence between the problem with
surface tension and our large $p$ growth, however
similar they appear to be, is rather indirect. The
issue of whether or not this growth model is indeed
characterized by an effective surface tension and how
the latter compares to the noise will be discussed
elsewhere.

\section{Anisotropic growth model}
\label{sec-anisotropy}

The iterated mapping model (\ref{F(f)}),
(\ref{bump-size}) can be generalized to describe
spatial anisotropy of local growth rate. Such an
anisotropy is characteristic of crystal growth, in
which all particles arriving at the crystal-liquid
interface have to accomodate to the anisotropic
crystal structure~\cite{rmpL80}.

Anisotropic growth oftenly gives rise to anisotropic
irregular fingering patterns called
dendrites~\cite{Vicsek84,paB86,Vicsek86,Eckmann1}. The
dynamics of dendrites growth obeys scaling laws
similar to that of the isotropic Laplacian
growth~\cite{praMBRS87,Couder90}. One of the
outstanding theoretical questions is how the scaling
exponents depend on the anisotropy.


In this problem, the cluster grows due to spatially
isotropic diffusion and aggregation of particles. Thus
the quasistationary probability distribution still
obeys Eq.~(\ref{bvp}). The new element, compared to
the isotropic model, is that the cluster change due to
particle freezing at the boundary depends on the local
growth direction $\vec v$, $|\vec v|=1$. (The unit
vector $\vec v$ is normal to the boundary.)
Accordingly, the probability of joining the cluster
becomes a function of $\vec v$, and Eq.~(\ref{dPdl})
is replaced by
  \be
dP=\Omega(\vec v)\,|\nabla u|\,dl\,, \quad
dl\subset {\cal D}_{n-1}\,.
  \ee
where the function $\Omega(\vec v)$ describes
anisotropy.

In order to include anisotropy in the mapping model
(\ref{F(f)}), (\ref{bump-size}), at the $n$-th growth
step one has to be able to predict local growth
direction ${\vec v}_n$ from particle positions
described by randomly chosen angles $\theta_k$, $k =
1, 2,..., n-1$. This is possible because
complex-valued Jacobian of a conformal mapping keeps
track of the angle change under the mapping.
Specifically, consider $\Theta_n = \theta_n + \arg
J_{n-1}$, where $J_{n-1}$ is given by
(\ref{Jacobian}). Then $\Theta_n$ defines a normal to
the cluster boundary, $\vec v_n=\cos\Theta_n\widehat
x+\sin\Theta_n\widehat y$, at the growth point
$F_{n-1}(e^{i\theta_n})$.

Now, there are several possible ways to account for
the growth anisotropy. For instance, 
one can introduce the
anisotropy by making $\lambda_n$ a function of
$\Theta_n$, {e.g.}, $\lambda_n \propto
\Omega^{1/2}(\vec v_n)$. Another way is to introduce
acceptance probability for the particles which depends
on $\Theta_n$ in some way. In the simulations reported
below we use an acceptance window for $M\Theta_n$ with
$M = 3,4,...$, corresponding to the growth with an
$M$-fold symmetry. Only particles with $\Theta_n$ such
that
  \be\label{theta-window}
-\theta_{\rm max} \le M \Theta_n \le \theta_{\rm max}
  \ee
are accepted. Here $\theta_{\rm max}$ is a parameter
in the interval $[0,\pi]$ controlling the degree of
anisotropy. Small values $\theta_{\rm max}\ll\pi$
correspond to highly anisotropic growth, whereas fully
isotropic growth is recovered in the limit
$\theta_{\rm max}\to\pi$.

Other aspects of the simulation are the same as in
Section~\ref{sec-nonoise}. We employed the elementary
mapping $\tilde f_{\lambda,p}(z)$ of the form
(\ref{new-tilde-f}) with the noise level controlled by
the parameter $p\ge1$.  Particles with large areas
were eliminated using the acceptance window
$[0,3a_\ast]$ defined in Section~\ref{sec-nonoise}.
An example of growth with the three-fold symmetry
($M = 3$) is shown in Fig.~\ref{p15a3}. In this case,
we used $\lambda_0 = 0.8$, $p = 1.5$, and $\theta_{\rm
max}=\cos^{-1}(0.9) \approx 0.451$. The cluster is
characterized by overall symmetric main branches
covered with numerous side branches.

In our model one has a separate control over the
degree of anisotropy and over noise, via the
parameters $\theta_{\rm max}$ and $p$.  This is
convenient for studying the effects of noise on the
ordering of branches in dendrites. To illustrate that,
we compare two growths with four-fold symmetry,
displayed in Figs.~\ref{p1a4},~\ref{p2a4}.  The
cluster in Fig.~\ref{p1a4} is obtained using the $p=1$
model without noise reduction, as described in
Sections~\ref{sec-model},~\ref{sec-areas}. The cluster
in Fig.~\ref{p2a4} is grown using the noise reduced
model of Section~\ref{sec-nonoise} with the parameter
$p=2$. In both cases, we use the same anisotropy
parameter: $\theta_{\rm max}=\cos^{-1}(0.95)\approx
0.318$. One notes high anisotropy of the growth
present at small scales in both cases, which is
significantly suppressed at larger scales for the
noisy growth with $p=1$ (see Fig.~\ref{p1a4}).
However, the $p=2$ growth with low noise remains very
anisotropic at all scales (see Fig.~\ref{p2a4}).

It is known from studies of on-lattice DLA models that
noise, no matter how strong, gives way to anisotropy
at sufficiently large
scales~\cite{jpaBB85,jpaM85,Vicsek86,Eckmann2}. We
thus expect that a similar effect may take place in
the noisy growth with $p=1$, making the growth shown
in Fig.~\ref{p1a4} at larger scales looking like that
in Fig.~\ref{p2a4}.

In agreement with the studies of the off-lattice DLA
models~\cite{paB86}, we observed that the dendritic
growth with the symmetry of order $M=3,\,4$ is much
more stable with respect to noise than that with
$M=5,\,6$ or higher. Scaling properties of the
anisotropic growth will be studied in
Section~\ref{sec-scaling}.

\begin{minipage}{3.1in}
\begin{figure}
  \begin{center}
    \begin{picture}(216,139)
      \epsfxsize=3in
      \put(0,-5){\epsfbox{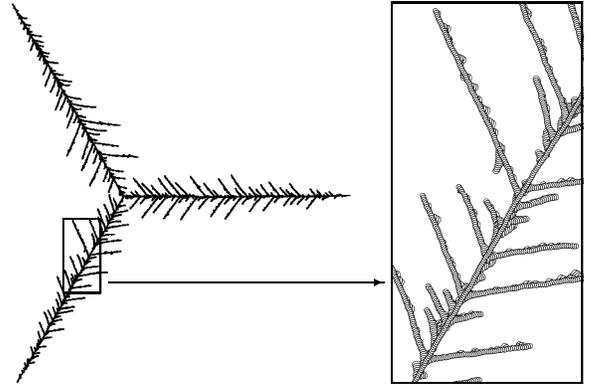}}
    \end{picture}
  \end{center}
  \caption{Anisotropic growth with $M=3$ obtained
     using the window (\protect\ref{theta-window}) for
     the growth direction with $\cos\theta_{\rm
     max}=0.9$. Other parameters used: $N = 10146$,
     $\lambda_0 = 0.8$, $p = 1.5$. \label{p15a3}}
\end{figure}
\vspace{-3mm}
\end{minipage}

\begin{minipage}{3.1in}
\begin{figure}
  \begin{center}
    \begin{picture}(216,136)
      \epsfxsize=3in
      \put(0,-5){\epsfbox{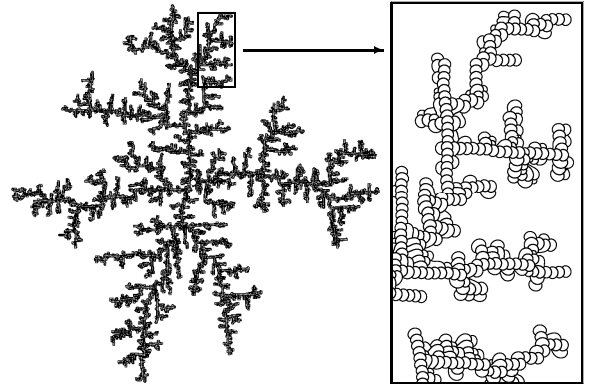}}
    \end{picture}
  \end{center}
  \caption{Anisotropic growth with $M = 4$, 
    $\cos\theta_{\rm max} = 0.95$, and high noise:
    $p=1$, $N = 7635$, $\lambda_0 = 0.3$.
    \label{p1a4}}
\end{figure}
\vspace{-3mm}
\end{minipage}

\begin{minipage}{3.1in}
\begin{figure}
  \begin{center}
    \begin{picture}(216,154)
      \epsfxsize=3in
      \put(0,-5){\epsfbox{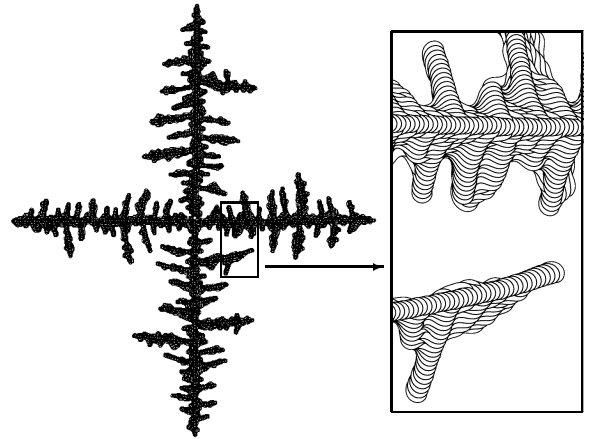}}
    \end{picture}
  \end{center}
  \caption{Anisotropic growth with $M = 4$,
    $\cos\theta_{\rm max} = 0.95$, and low noise: $p=2$,
    $N = 6782$, $\lambda_0 = 0.8$. \label{p2a4}}
\end{figure}
\vspace{0mm}
\end{minipage}

\section{Scaling properties}
\label{sec-scaling}

Scaling of $R_n$ for all growth models introduced
above is studied here using the following procedure.
The cluster radius $R_n$ obtained from
(\ref{R-definition}), (\ref{R-p=1}), (\ref{new-f'}),
is plotted against the cluster area $A_n$, evaluated
as a sum of individual particle areas $a_n$.  
Asymptotically, at large $n$, one has $R_n\propto
A_n^{1/d}$. To determine $d$ more accurately we
optimize initial conditions of the growth, represented
in our model by nondimensionalized particle size
$\lambda_0$, as described below.

In the log--log plot of $R_n$ versus $A_n$ one can
clearly distinguish two regimes, initial growth and
developed, or {\it regular} growth, characterized by
somewhat different slopes of the corresponding parts
of the $\ln R$~vs.~$\ln A$ curves. Geometrical
meaning of these regimes is as follows. For isotropic
growth, with or without noise suppression, the cluster
initially consists of branches growing
practically independently. Later, at the regular
growth stage, the number of main branches is reduced
to four or five, all interacting and competing with
each other.  For anisotropic growth with the $M$-fold
symmetry the number of main branches is $M$ at all
stages of growth. Regular growth in this case is
distinguished by many fingers appearing on the sides
of $M$ main branches.

The initial stage is more pronounced when the particle
size, determined by the value of $\lambda_0$, is much
smaller than the unit circle from which the growth
starts. Since we are interested in the regular growth
scaling, in each case studied we tried to optimize the
value of $\lambda_0$ to shorten the initial growth
stage, carefully checking that the variation of
$\lambda_0$ has no detectable effect on the asymptotic
slope of the $\ln R$~vs.~$\ln A$ curve. The benefit of
shortening the initial growth stage is that, at
constant number of particles, it leads to longer
regular growth and thus allows to extract the scaling
exponent with higher precision. The resulting curves
are presented in Fig.~\ref{picF1}, as described in the
figure caption and below.

The optimal value of $\lambda_0$ determined for the
isotropic growth with $p=1$ is close to
$\lambda_0=0.8$.  For the scaling analysis we used the
growth displayed in Fig.~\ref{fig-p=1}, in which $N =
17545$, $\lambda_0 = 0.8$, $p = 1$.  In
Fig.~\ref{picF1}, it corresponds to the lowest of the
curves marked by ${\rm a}_1$. To eliminate the effect
of fluctuations at the initial stage of the growth, we
also generated the curves ${\rm a}_2$, ${\rm a}_3$,
and ${\rm a}_4$, by averaging $\ln R$ over 5, 10, and
50 growth realizations with $N=5000$, $1000$, and
$200$ time steps, respectively.

For isotropic growth with reduced noise we analyzed
two growths with $p=3$:

\centerline{curve ${\rm b}_1$ with $N = 11611$,
$\lambda_0 = 0.1$;}

\centerline{curve ${\rm b}_2$ with $N = 15043$,
$\lambda_0 = 0.2$.}

\noindent The curve ${\rm b}_2$ corresponds to the
growth displayed in Fig.~\ref{p3i}. At low noise, the
fluctuations of $R_n$ are quite small, which makes
additional averaging over realizations unnecessary.

Note that for the isotropic growth models the strategy
of optimizing $\lambda_0$ works quite well, allowing
to almost entirely eliminate the initial growth
region. The scaling dimension found from the slope of
the best straight line fits is close to $1.7$. To
study the deviation from $1.7$, we subtract from all
curves the linear function $\ln R=\ln A/1.7$ and plot
the result in the lower part of Fig.~\ref{picF1}.  
Note that upon this subtraction the curves for
isotropic growth, with or without noise suppression,
become nearly perfectly horizontal. Estimate of the
deviation from the best horizontal line fit shows that
the value $1.7$ is accurate within $1\%$.

\begin{minipage}{3.1in}
\begin{figure}
  \begin{center}
    \begin{picture}(216,320)
      \epsfxsize=3in
      \put(0,160){\epsfbox{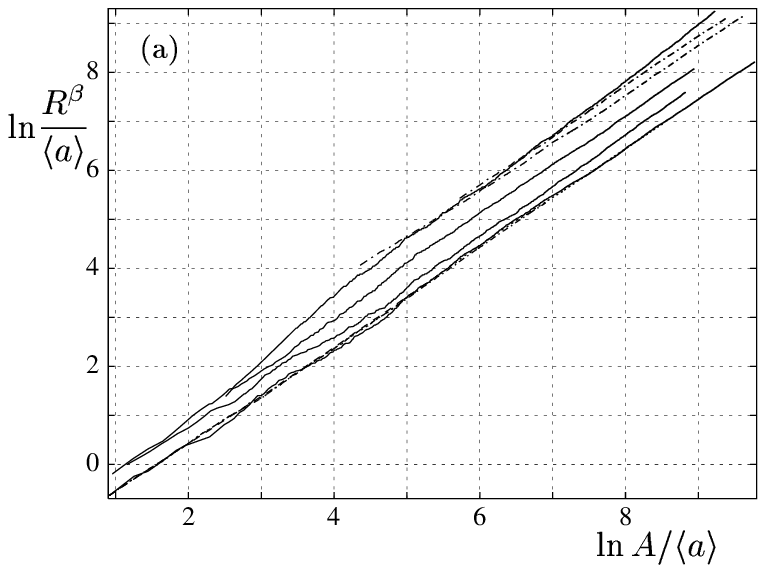}}
      \put(0,-5){\epsfbox{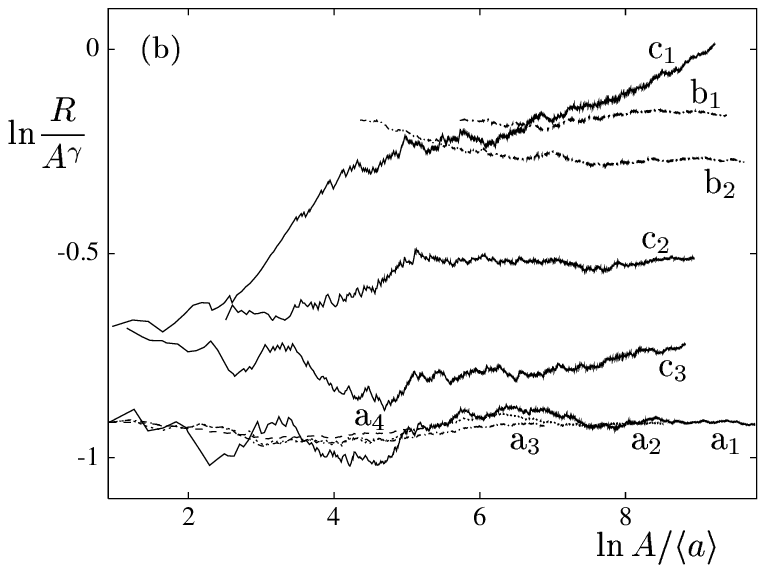}}
    \end{picture}
  \end{center}
  \caption{Log--log plots of $R^\beta/\langle a
    \rangle$ (a) and $R/A^\gamma$ (b) {\it vs.}
    normalized area $A/\langle a\rangle$ for several
    clusters described in the text. Here $\beta =
    1.7$, $\gamma = 1/1.7 = 1/\beta$, and $\langle
    a \rangle$ is the particle area averaged over the
    cluster. The plots (a) and (b) are connected by an
    affine transformation. \label{picF1}}
\end{figure}
\vspace{0mm}
\end{minipage}

For anisotropic models, we consider three different
growths:

\centerline{curve ${\rm c}_1$ with $N =10146$,
$\lambda_0 = 0.8$, $p = 1.5$, $M = 3$;}

\centerline{curve ${\rm c}_2$ with $N =7635$,
$\lambda_0 = 0.3$, $p = 1$, $M = 4$;}

\centerline{curve ${\rm c}_3$ with $N = 6782$,
$\lambda_0 = 0.8$, $p = 2$, $M = 4$.}

\noindent These curves correspond to the growths
displayed in Figs.~\ref{p15a3}, \ref{p1a4}, and
\ref{p2a4}, respectively. As above, we subtract 
the linear function $\ln
R=\ln A/1.7$. However, after this subtraction, the
curves ${\rm c}_1$ and ${\rm c}_3$ retain some
residual slope. Estimating it, we conclude that the
best value for the fractal dimension is $d \approx
1.5$ for ${\rm c}_1$, and $d \approx 1.62$ for ${\rm
c}_3$. The latter value agrees with the values
$d\approx 1.58$ and $d\approx 1.63$ of the growth with
$M=4$ reported in Refs.~\cite{Couder90,praMBRS87}.

For the curve ${\rm c}_2$ corresponding to anisotropic
growth with noise, after subtracting $\ln R=\ln A/1.7$
we do not find any significant residual slope. It is
possible, however, that the dimension $1.7$
corresponds to the crossover regime and changes to a
lower value at larger $N$. Similar behavior is known
to take place in the on-lattice DLA
growth~\cite{praMBRS87}, where the dimension $1.7$
observed for not very large clusters crosses over to
$1.63$ at $N\sim 4\cdot 10^6$.

To understand possible sources of errors in
determining the fractal dimension from $\ln
R$~vs.~$\ln A$ curves, here we consider how $R_n$ and
$A_n$ fluctuate with $n$. The fluctuations of $\ln
R_n$ gradually decrease with increasing $n$, as can be
clearly seen in the lower panel of Fig.~\ref{picF1}.  
A convenient way to analyze fluctuations is to plot
pairs $(\ln R_N, \ln A_N)$ for particular $N$,
repeating growth many times. In Fig.~\ref{clouds} we
present results for $10^3$ growth samples and several
values of $N$. The resulting clouds become more
compact as $N$ increases, indicating that the
fluctuations of $\ln R_N$ and $\ln A_N$ are
decreasing.

\begin{minipage}{3.1in}
\begin{figure}
  \begin{center}
    \begin{picture}(216,154)
      \epsfxsize=3in
      \put(0,-5){\epsfbox{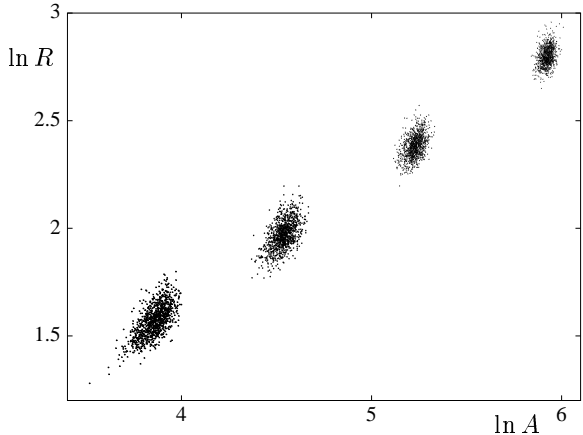}}
    \end{picture}
  \end{center}
  \caption{Clouds of points $(\ln R, \ln A)$
    corresponding to $1000$ realizations for $N = 25$,
    $50$, $100$, and $200$, $\lambda_0 = 0.8$, $p =
    1$. \label{clouds}}
\end{figure}
\vspace{-1mm}
\end{minipage}

Let us first discuss fluctuations of $\ln A_n$.  The
total area $A_n$ is the sum of individual particles
areas $a_k$, $k = 1, 2, ..., n$. Assuming that the
areas $a_k$ are independent or, more precisely, have
only short correlations, one obtains a Gaussian
distribution with the variance $\propto n$. (As we
argue below, there exist long negative correlation of
particle areas, which may further reduce fluctuations
of $A_n$.) The fluctuations of $\ln A_n$ are simply
given by relative fluctuations $\delta A_n/A_n$, which
means that for large $n$ the distribution of $\ln A_n$
is also Gaussian, with the variance $\propto n^{-1}$.

On the other hand, the radius $R_n$ is {\it a product}
(\ref{R-definition}) of stretching factors
$J_k^{(\infty)}=f_k'(z \to \infty)$.  Since
$J_k^{(\infty)}>1$ for all $k$, the quantity $R_n$
grows monotonically, so that $R_n \propto A_n^{1/d}$
at large $n$. Thus the noise due to fluctuations of
$J_k^{(\infty)}$ is of a multiplicative nature. One
can write
  \be\label{log-sum}
\ln R_n = \sum_{k=1}^n\, \ln f_k'(z \to \infty)
\,,
  \ee
which suggests that the distribution of $\ln R_n$ is
Gaussian, i.e., the distribution of $R_n$ is
log-normal. Indeed, the log-normal fit perfectly
describes the statistics of $R_n$, as demonstrated in
Fig.~\ref{Rpdf}. However, attempting a Gaussian fit
produces an asymmetric distribution deviating from the
observed distribution of $\ln R_n$. Thus, even though
the relative fluctuations of $R_n$ are small, the
statistics is best described as log-normal.

\begin{minipage}{3.1in}
\begin{figure}
  \begin{center}
    \begin{picture}(216,155)
      \epsfxsize=3in
      \put(0,-5){\epsfbox{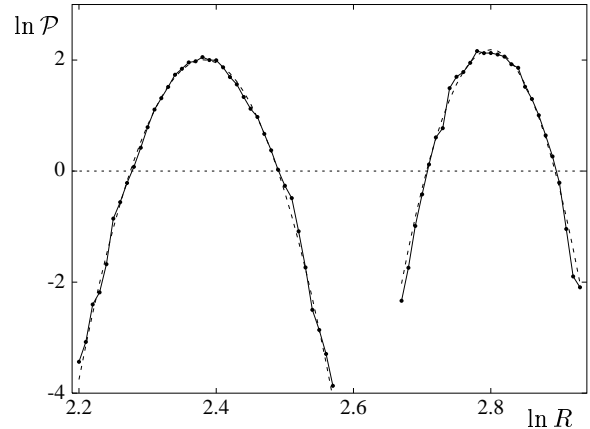}}
    \end{picture}
  \end{center}
  \caption{Probability distribution of $\ln R$
    for $N = 100$ and $N = 200$ calculated
    from $10368$ and $2862$ growth
    realizations respectively, $\lambda_0 = 0.8$, $p =
    1$. \label{Rpdf}}
\end{figure}
\vspace{0mm}
\end{minipage}

Naively, Eq.~(\ref{log-sum}) implies growth of the
variance of $\ln R_n$ with $n$. However,
Figs.~\ref{clouds},~\ref{Rpdf} demonstrate that, on
the contrary, the width of $\ln R_n$ distribution is
decreasing with $n$. This nontrivial behavior was
first mentioned, without explanation, in
Ref.~\cite{P1}.

To rationalize the observed sharpening of the
distribution of $R_n$, one can argue as follows. We
note that the dynamics of $R_n$ is characterized by a
negative feedback. Consider growth of a cluster which
at the $n$-th step has a radius smaller than average.
Then the Jacobian of $F_n$ is typically smaller than
its mean value at this number of particles. In this
case, according to (\ref{bump-size}), subsequently
growing particles will have larger $\lambda_k$'s, and
thus larger areas, until the cluster radius will
approach the average value. The evolution of a cluster
which at a certain step has a radius larger than
average can be considered in a similar way. This
long-time anticorrelation of $\lambda_k$'s suppresses
the fluctuations of $R_n$. Also, it produces long
negative correlation of particle areas.

\section{Summary}


To conclude, growth models using conformal mappings
have large flexibility allowing for independent
control over noise and growth anisotropy. We
generalized the model~\cite{HL} by using flat
particles to suppress noise. It is essential that
these models lead to an intrinsically isotropic growth
with reduced noise, in contrast with other previously
studied models. Also, we demostrated that favoring
growth in certain directions can be used to simulate
anisotropy of the growth rate.

Having separate control on the noise and anisotropy,
we have been able to analyze their effect on scaling
properties. We found that the fractal dimension
$d=1.7$, universally for any isotropic growth,
regardless of the noise level. However, the fractal
dimension is somewhat reduced in the presence of
anisotropy.

It was assumed~\cite{HL} that particle size
fluctuations, present in the conformal mapping model,
are insignificant. We observed that the growth rules
used in Ref.~\cite{HL} lead to occasional appearance
of exceptionally large particles. We have shown that
by augmenting the model with an area acceptance
criterion this problem is fixed.

Clearly, more work has to be done to establish a
relation of the introduced models with real physical
processes, like viscous fingering or dendritic crystal
growth. Another interesting open question is how to
introduce an effective surface tension.

\acknowledgments

We thank G.E.~Falkovich, M.B.~Hastings, V.A.~Kazakov,
V.V.~Lebedev, B.Z.~Spivak, and P.B.~Wiegmann for
useful discussions.  Hospitality of Weizmann Institute
made our collaboration possible. This work was
partially supported by the Minerva Foundation (M.S.)
and by the NSF Award 67436000IRG (L.L.).

\appendix

\section{Insides of the particle area distribution}
\label{appen-areas}

Here we discuss in more detail the distribution of
particle areas. The main feature manifest in the area
histogram plotted in Fig.~\ref{arpdf} is a sharp
asymmetric peak at $\approx 2.1 \, a_\ast$. This peak
has its origin in the dependence of particle size on
the growth point.

The argument is as follows. First we note that the
growth is taking place predominantly at the tips of
the branches. Because of that, for several particles
growing on each other, there is a tendency to preserve
growth direction. This leads to formation of
relatively long chains of particles growing in a
particular direction, clearly seen in the inset of
Fig.~\ref{fig-p=1}. The chains are mostly formed at
the tips of outer branches.

Now, consider a particle growing near one of the tips.
The area of this particle has some dependence on the
position of the growth point relative to the tip. The
peak in the histogram in Fig.~\ref{arpdf} is explained
if one assumes that the particle area has a local
minimum in the forward growth direction. The minimum
in the area leads to a caustic in the histogram.
Ideally, this would have produced an asymmetric square
root singularity with probability equal zero on the
left side. Because of particle size variation among
branches, the singularity is smeared into a peak.

\begin{minipage}{3.1in}
\begin{figure}
  \begin{center}
    \begin{picture}(216,312)
      \epsfxsize=3in
      \put(0,157){\epsfbox{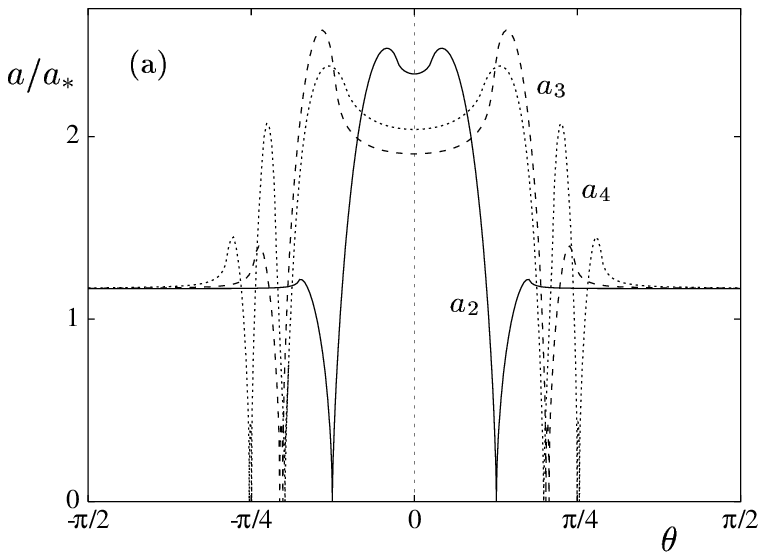}}
      \put(0,-5){\epsfbox{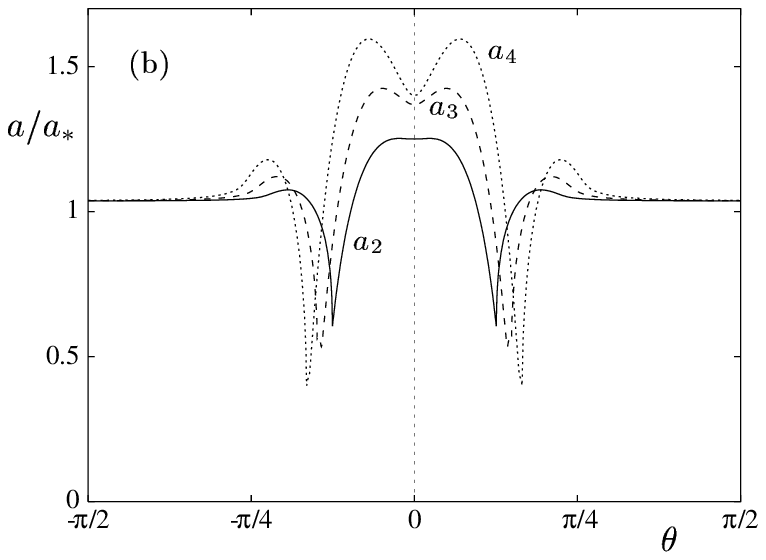}}
    \end{picture}
  \end{center}
  \caption{Area of a particle as a function of its
    growth point, characterized by $\theta$.
    \label{fig-a-theta}}
\end{figure}
\vspace{0mm}
\end{minipage}

To verify the above assumption, we consider areas for
the first few particles grown on the $|z| = 1$ circle
with the parameter $\lambda_0=0.2$. The area of the
very first particle is close to $1.2 \, a_\ast$ and,
according to (\ref{a-star}), is independent of its
position. The area of the second particle $a_2$
depends on its position $\theta$ relative to the first
particle, as shown by a solid line in
Fig.~\ref{fig-a-theta}(a). Note that the area is the
same as that of the first particle when the particles
are far apart, $\theta\gg\lambda_0$, and is overall
substantially larger when the particle overlap,
$\theta\sim\lambda_0$. Partially, this is explained by
the dependence (\ref{a-star}) of particle area on the
circle radius. (Assuming that $a_2(\theta \sim
\lambda_0)$ can be crudely estimated by (\ref{a-star})
with $r=\lambda_0$.) Another effect that contributes
to the area $a_2$ increase for $\theta\sim\lambda_0$
is the variation of the Jacobian as a function of
$\theta$, leading to additional stretching of the
second particle.

The feature in Fig.~\ref{fig-a-theta}(a) which is of
interest in connection to the peak in the area
distribution ${\cal P}(a)$ is the minimum of
$a_2(\theta)$ at $\theta=0$. Translated to the
histogram of areas, it leads to a caustic described by
a square root singularity. However, as a possible
explanation of the peak in Fig.~\ref{arpdf} this is
only partially satisfying, since one has to understand
why similar caustics due to the two maxima of
$a_2(\theta)$ are not observed in Fig.~\ref{arpdf}.

The reason for the difference between the effects of
maxima and minima can be seen from a comparison with
the case of three and four particles.  Consider the
situation when the second particle is centered exactly
on the first particle, and the third particle is grown
at an angular position $\theta$ relative to the first
two particles. The area of the third particle
$a_3(\theta)$ is plotted in Fig.~\ref{fig-a-theta}(a)
in a dashed line. Note that, since the curvature at
the minimum of $a_3(\theta)$ at $\theta=0$ is much
smaller than for $a_2(\theta)$, the corresponding
caustic in ${\cal P}(a)$ will be much stronger. On the
other hand, the curvature at the maxima of
$a_3(\theta)$ is about the same as that for
$a_2(\theta)$. Both observations remain correct for
any number of particles. To illustrate this we plot
the area $a_4(\theta)$ of the fourth particle in the
presence of three particles grown exactly on the top
of each other --- see dotted line in
Fig.~\ref{fig-a-theta}(a).

Another notable feature in the plots of
$a_{2,3,4}(\theta)$ is that the area becomes much
smaller than $a_\ast$, approaching zero near certain
values of $\theta$. This behavior is related to the
growth near particle corners, which are the points of
divergence of the Jacobian. According to
(\ref{bump-size}), larger Jacobian translates into
smaller particle area. The particles growing near
corners form the tail of the area distribution ${\cal
P}(a)$ at small areas $a\ll a_\ast$. The behavior of
${\cal P}(a)$ in this tail, ${\cal P}(a)\propto
a^{1/2}$, follows from the square root divergence of
the Jacobian at particle corners. The slope $1/2$ is
clearly seen in the $\ln{\cal P}$~vs.~$\ln a$ plot in
Fig.~\ref{artail}.

The features in $a_{2,3,4}(\theta)$ discussed above
evolve in an interesting way for the models with lower
noise corresponding to $p>1$ ---~see
Fig.~\ref{fig-a-theta}(b). The plots of
$a_{2,3,4}(\theta)$ in this figure are produced for
the model with $\lambda_0=0.2$ and $p=3$ in the same
way as above for $p=1$. Note that relative changes of
the area as a function of $\theta$ are smaller than
for $p=1$. One reason for this is in weaker
curvature variation for flat particles, which makes
particle area less sensitive to the growth point
position. Another reason is that at $p>1$ the
particles corners have no cusps, and thus particles
with small areas do not appear.

\section{Discussion of the numerical method}
\label{appen-numerics}

Here we comment on the optimal choice of the numerical
procedure. First, since the areas of new particles are
evaluated before the particles are accepted, one
could, instead of eliminating large particles, change
the growth algorithm so that all particles areas
become equal. This can be achieved by adjusting the
parameter $\lambda_n$ for each particle until its area
converges to a given value. Although this would
somewhat slow down the speed of computation, an
obvious gain would be in a more direct relation with
the standard DLA growth.

Also, one could attempt at increasing the speed and
efficiency of the growth algorithm by introducing in
it various improvements:

(i) Coarsening of the mappings which correspond to
particles sufficiently deep in the stagnation regions.
It was demonstrated in Ref.~\cite{HL} that an accurate
envelope of the cluster can be obtained by using
truncated Laurent series of $F_n(z)$. One can
implement this observation as follows. At the growth
step $n$ choose some $1 < m < n$ in such a way that
all particles with the numbers $\le m$ are located
sufficiently deep inside the stagnation part of the
cluster. Then one can replace the mapping $F_n = f_1
\circ ... \circ f_n$ by
  \be
F_n^{\rm (approx)} = \left[f_1\circ ...\circ
f_m\right]_{\rm truncated}
\circ f_{m+1}\circ ...\circ f_n
\,,
  \ee
where the mapping in parentheses which is replaced by
truncated series is nothing but $F_m$. One can choose
$m$ so that the finite series representation of the
mapping $F_m(z)$ is accurate for $z$ in the active
growth region. By this trick, instead of computing a
composition of $n$ functions, one has to deal with
only $n-m$ functions at each growth step. Since at
large $n$ most of the particles are in the stagnation
regions, one may have $n-m \ll n$.

(ii) Evaluating the particle area with lower
precision. We used several hundred points on 
each particle's boundary, which produces areas
accurate within $0.1\%$. Practically, such a high
precision may not be necessary. Instead, one can
predict particle areas by estimating the Jacobian at
several points chosen within the bump according to
some rule or randomly.

(iii) Using an area acceptance window to discriminate
against very small particles with areas $\ll a_\ast$.
These particles practically do not change the
structure of the cluster branches, except near the
corners between adjacent particles. However, due to
the presence of small particles additional mappings
appear in the composition sequence $f_n\circ ...\circ
f_1$, which slows down the computation.

We have not used these procedures in the simulations
described above and neither systematically studied
their efficiency. We felt that, at the initial stage,
keeping growth algorithm as precise and simple as
possible, even at the price of somewhat slowing it
down, makes the results more solid.

\end{multicols}

\end{document}